\documentclass[pre,twocolumn,superscriptaddress]{revtex4-1}

\usepackage{graphicx}
\usepackage{amsmath}  
\usepackage{esint} 
\usepackage{color} 

\usepackage{dcolumn}
\usepackage{bm}

\usepackage[normalem]{ulem}
\useunder{\uline}{\ul}{}

\begin{document}

\title{Copula-based algorithm for generating bursty time series}

\author{Hang-Hyun Jo}
\email{hang-hyun.jo@apctp.org}
\affiliation{Asia Pacific Center for Theoretical Physics, Pohang 37673, Republic of Korea}
\affiliation{Department of Physics, Pohang University of Science and Technology, Pohang 37673, Republic of Korea}
\affiliation{Department of Computer Science, Aalto University, Espoo FI-00076, Finland}
\author{Byoung-Hwa Lee}
\affiliation{Department of Physics, Pohang University of Science and Technology, Pohang 37673, Republic of Korea}
\affiliation{Asia Pacific Center for Theoretical Physics, Pohang 37673, Republic of Korea}
\author{Takayuki Hiraoka}
\affiliation{Asia Pacific Center for Theoretical Physics, Pohang 37673, Republic of Korea}
\author{Woo-Sung Jung}
\email{wsjung@postech.ac.kr}
\affiliation{Department of Physics, Pohang University of Science and Technology, Pohang 37673, Republic of Korea}
\affiliation{Department of Industrial and Management Engineering, Pohang University of Science and Technology, Pohang 37673, Republic of Korea}
\affiliation{Asia Pacific Center for Theoretical Physics, Pohang 37673, Republic of Korea}

\date{\today}

\begin{abstract}
    Dynamical processes in various natural and social phenomena have been described by a series of events or event sequences showing non-Poissonian, bursty temporal patterns. Temporal correlations in such bursty time series can be understood not only by heterogeneous interevent times (IETs) but also by correlations between IETs. Modeling and simulating various dynamical processes requires us to generate event sequences with a heavy-tailed IET distribution and memory effects between IETs. For this, we propose a Farlie-Gumbel-Morgenstern copula-based algorithm for generating event sequences with correlated IETs when the IET distribution and the memory coefficient between two consecutive IETs are given. We successfully apply our algorithm to the cases with heavy-tailed IET distributions. We also compare our algorithm to the existing shuffling method to find that our algorithm outperforms the shuffling method for some cases. Our copula-based algorithm is expected to be used for more realistic modeling of various dynamical processes.
\end{abstract}

\maketitle

\section{Introduction}

Dynamical processes in various natural and social phenomena have been described by a series of events or event sequences showing non-Poissonian, bursty temporal patterns. Examples include solar flares~\cite{Wheatland1998WaitingTime}, earthquakes~\cite{Corral2004LongTerm, deArcangelis2006Universality}, neuronal firings~\cite{Kemuriyama2010Powerlaw}, and human activities~\cite{Barabasi2005Origin, Karsai2018Bursty}. Temporal correlations in such bursty time series can be understood not only by heterogeneous interevent times (IETs) but also by correlations between IETs~\cite{Goh2008Burstiness, Jo2017Modeling}. Here the IET, denoted by $\tau$, is defined as a time interval between two consecutive events. On the one hand, heterogeneities of IETs have been characterized by heavy-tailed or power-law IET distributions $P(\tau)$ with a power-law exponent $\alpha$~\cite{Karsai2018Bursty}:
\begin{equation}
    \label{eq:Ptau_simple}
    P(\tau)\sim \tau^{-\alpha}.
\end{equation}
On the other hand, the correlations between IETs can be characterized by a memory coefficient~\cite{Goh2008Burstiness} among other measures such as local variation~\cite{Shinomoto2003Differences} and a bursty train~\cite{Karsai2012Universal}. The memory coefficient $M$ is defined as the Pearson correlation coefficient between two consecutive IETs, whose value for a sequence of $n$ IETs, i.e., $\{\tau_i\}_{i=1,\cdots,n}$, can be estimated by
\begin{equation}
    M \equiv\frac{1}{n - 1}\sum_{i=1}^{n-1}\frac{(\tau_i - \mu_1)(\tau_{i+1} - \mu_2)}{\sigma_1 \sigma_2},
    \label{eq:memory_original}
\end{equation}
where $\mu_1$ ($\mu_2$) and $\sigma_1$ ($\sigma_2$) are the average and the standard deviation of the first (last) $n-1$ IETs, respectively. Positive $M$ implies that the small (large) IETs tend to be followed by small (large) IETs. Negative $M$ implies the opposite tendency, while $M=0$ is for the uncorrelated IETs. We mainly focus on the case with positive $M$, based on the empirical observations~\cite{Goh2008Burstiness, Wang2015Temporal, Guo2017Bounds, Bottcher2017Temporal}. 

The dynamical processes, such as spreading and diffusion, taking place in a network of individuals are known to be strongly affected by bursty interaction patterns between individuals~\cite{Vazquez2007Impact, Karsai2011Small, Miritello2011Dynamical, Rocha2011Simulated, Jo2014Analytically, Perotti2014Temporal, Delvenne2015Diffusion, Artime2017Dynamics, Hiraoka2018Correlated}. This topic has been studied in a framework of temporal networks~\cite{Holme2012Temporal, Masuda2016Guide}, where a link connecting two nodes is considered being existent or activated only at the moment of interaction. For example, an infectious disease, information, or a random walker on a node, say $A$, can be transferred to the $A$'s neighboring node, say $B$, only when $A$ interacts with $B$. Therefore, the dynamical behavior, such as spreading speed, would be influenced by how often these nodes interact with each other, more generally, by the temporal interaction pattern between them. This temporal interaction pattern can be described by the event sequence when each event denotes the interaction. Consequently, modeling and simulating realistic temporal networks often requires us to generate event sequences for links (or nodes) showing the empirically observed properties, such as heavy-tailed IET distribution in Eq.~\eqref{eq:Ptau_simple} and/or memory effects measured by the memory coefficient in Eq.~\eqref{eq:memory_original}.

It is straightforward to generate event sequences characterized only by the IET distribution $P(\tau)$, i.e., when IETs are fully uncorrelated with each other as in the renewal process~\cite{Mainardi2007Poisson}: $n$ IETs are independently drawn from the given $P(\tau)$ to make a sequence of IETs, $\{\tau_i\}_{i=1,\cdots,n}$. Provided that the first event occurs at time $t_0$, the events are considered to occur at times $t_i=t_0+\sum_{i'=1}^i \tau_{i'}$ for $i=1,\cdots,n$.

In contrast, the generative methods for event sequences with correlated IETs have not been fully explored, except for a few recent works. We focus on two existing methods: The first method is to shuffle a set of IETs, prepared using a given $P(\tau)$, to implement the desired value of $M$ in Eq.~\eqref{eq:memory_original}~\cite{Hiraoka2018Correlated}. By this method one can control $P(\tau)$ and $M$ independently, while the shape of $P(\tau)$ may set bounds on the range of $M$~\cite{Guo2017Bounds}. Note that for the implementation of this method, the number of IETs should be predetermined, which however is not necessarily the case. For example, let us consider the simulation of a dynamical process using the event sequence; it is often the case that how many IETs are to be needed for reaching a stationary state of the process may not be known a priori. The second method is to generate IETs sequentially using the conditional probability distribution $P(\tau|\tau')$~\cite{Artime2017Dynamics}, by which a next IET $\tau$ is drawn given the previous IET $\tau'$. One can generate an arbitrary number of IETs without predetermining the number of IETs. However, since the function used in Ref.~\cite{Artime2017Dynamics} is $P(\tau|\tau')=\alpha (1+\tau')^{\alpha}/(\tau+\tau')^{\alpha+1}$, $\alpha$ controls both the heterogeneity of IETs and the degree of memory effects, implying the inevitable dependency between the statistics of IETs and correlations between IETs. 

To overcome disadvantages in the previous methods, we propose an alternative generative method using the conditional probability distribution $P(\tau|\tau')$ but based on the Farlie-Gumbel-Morgenstern copula~\cite{Nelsen2006Introduction, Takeuchi2010Constructing}, by which one can generate an arbitrary number of IETs without predetermining the number of IETs, while the statistics of IETs and their correlations can be independently controlled. In Sec.~\ref{sec:algo} we introduce the copula-based algorithm for generating event sequences with correlated IETs. In Sec.~\ref{sec:result} we apply our algorithm to the cases with exponential and power-law IET distributions as well as power-law IET distribution with exponential cutoff, and we also discuss the performance of our algorithm for different cases in terms of computational times, in comparison to the shuffling method used in Ref.~\cite{Hiraoka2018Correlated}. Finally, we conclude our work in Sec.~\ref{sec:conclusion}.

\section{Copula-based algorithm}\label{sec:algo}

We introduce the copula-based algorithm for generating event sequences with correlated interevent times (IETs) for a given IET distribution $P(\tau)$ and memory coefficient $M$ between two consecutive IETs. For this, we model the joint probability distribution $P(\tau_i,\tau_{i+1})$ by adopting a Farlie-Gumbel-Morgenstern (FGM) copula among others~\cite{Takeuchi2010Constructing, Nelsen2006Introduction}. It is because the FGM copula is simple and analytically tractable, despite the range of correlation being somewhat limited, which will be discussed below. The joint probability distribution based on the FGM copula is written as~\footnote{The FGM copula is a function $C$ joining a bivariate cumulative distribution function (CDF) to their one-dimensional marginal CDFs such that $G(x,y)=C[u(x),v(y)]=uv[1+r(1-u)(1-v)]$, where $u$ and $v$ are CDFs of variables $x$ and $y$, respectively~\cite{Takeuchi2010Constructing, Nelsen2006Introduction}. Then the bivariate probability distribution function (PDF) of $x$ and $y$ is obtained by $\frac{\partial^2G}{\partial x\partial y}=P_1(x)P_2(y)[1+r(2u-1)(2v-1)]$, where $P_1(x)$ and $P_2(y)$ denote PDFs of $x$ and $y$, respectively. The FGM copula for IETs has recently been found to be useful for the analytical approach to autocorrelation functions~\cite{Jo2019Analytically}.}
\begin{equation}
	P(\tau_i,\tau_{i+1}) = P(\tau_i) P(\tau_{i+1}) \left[1+r f(\tau_i)f(\tau_{i+1})\right],
	\label{eq:Ptautau}
\end{equation}
where 
\begin{equation}
    f(\tau) \equiv 2F(\tau)-1,\ F(\tau) \equiv \int_0^{\tau} d\tau' P(\tau').
\end{equation}
Here $F(\tau)$ denotes the cumulative distribution function (CDF) of $P(\tau)$, and $P(\tau_i)$ and $P(\tau_{i+1})$ are assumed to have the same functional form. The parameter $r$, controlling the correlation between two consecutive IETs, is in the range of $-1\leq r\leq 1$ because $P(\tau_i,\tau_{i+1})\geq 0$ and $0\leq F(\tau)\leq 1$, hence $-1\leq f(\tau)\leq 1$. It is straightforward to relate $r$ with the memory coefficient $M$ in Eq.~\eqref{eq:memory_original} using the FGM copula in Eq.~\eqref{eq:Ptautau} as follows:
\begin{equation}
    M \simeq\frac{ \langle \tau_i\tau_{i+1}\rangle - \mu^2 }{\sigma^2 } = \frac{r}{\sigma^2} \left[ \int_0^\infty d\tau \tau P(\tau)f(\tau) \right]^2 \equiv ar,
    \label{eq:Mr_ratio}
\end{equation}
where 
\begin{equation}
    \langle \tau_i\tau_{i+1} \rangle \equiv \int_0^\infty d\tau_i \int_0^\infty d\tau_{i+1} \tau_i\tau_{i+1} P(\tau_i,\tau_{i+1}),
\end{equation}
and $\mu$ and $\sigma$ are the mean and standard deviation of IETs, respectively. The positive constant $a$ is determined only by $P(\tau)$, irrespective of $M$. The upper bound of $a$ is $1/3$ for any $P(\tau)$ as proven in Ref.~\cite{Schucany1978Correlation}, i.e., $|M|\leq a= 1/3$. Despite such relatively weak correlations by the FGM copula, one can still study the FGM copula as the empirical values of $|M|$ tend to be relatively small in several cases~\cite{Goh2008Burstiness, Wang2015Temporal, Guo2017Bounds}.

\begin{figure}[!t]
    \includegraphics[width=\columnwidth]{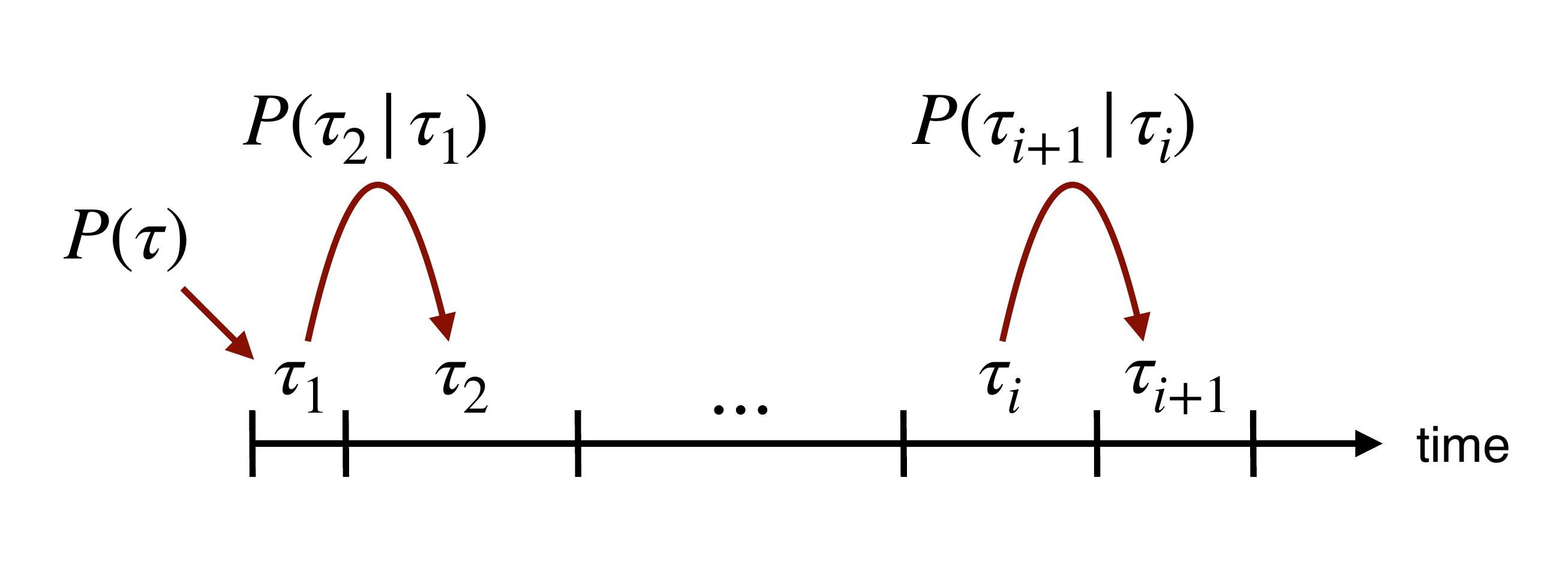}
    \caption{Schematic diagram of the copula-based algorithm using a given interevent time (IET) distribution $P(\tau)$ and a conditional probability distribution $P(\tau_{i+1}|\tau_i)$ for two consecutive IETs in Eq.~\eqref{eq:cond_P}.}
    \label{fig:diagram}
\end{figure}

Using the joint probability distribution in Eq.~\eqref{eq:Ptautau}, one can get the conditional probability distribution of $\tau_{i+1}$ for a given $\tau_i$:
\begin{equation}
    \label{eq:cond_P}
    P(\tau_{i+1}|\tau_{i})=\frac{P(\tau_{i},\tau_{i+1})}{P(\tau_{i})}=P(\tau_{i+1}) \left[ 1+ r f(\tau_{i})f(\tau_{i+1}) \right].
\end{equation}
In order to draw a value of $\tau_{i+1}$ from $P(\tau_{i+1}|\tau_i)$, we use the inverse transform sampling or transformation method~\cite{Clauset2009Powerlaw}: For a random number $x$ drawn from a uniform distribution defined in $0\leq x< 1$, we obtain $\tau_{i+1}$ for a given $\tau_i$ by solving the equation
\begin{equation}
    1-x = F(\tau_{i+1}|\tau_i),
    \label{eq:tau_next}
\end{equation}
where
\begin{equation}
    F(\tau_{i+1}|\tau_{i}) \equiv \int_{0}^{\tau_{i+1}}d\tau' P(\tau'|\tau_{i}).
    \label{eq:cond_F}
\end{equation}
For convenience, we define 
\begin{equation}
    c_i\equiv rf(\tau_i)=r[2F(\tau_i)-1],
\end{equation}
to rewrite Eq.~\eqref{eq:cond_F} as
\begin{eqnarray}
    && F(\tau_{i+1}|\tau_{i})\nonumber\\
    &&= (1-c_i) \int_{0}^{\tau_{i+1}}d\tau' P(\tau') + 2c_i \int_{0}^{\tau_{i+1}}d\tau' P(\tau') F(\tau')\nonumber\\
    &&= (1-c_i) F(\tau_{i+1}) + c_i F(\tau_{i+1})^2.
\end{eqnarray}
Then from Eq.~\eqref{eq:tau_next} with $y\equiv F(\tau_{i+1})$, one gets
\begin{equation}
    c_i y^2 + (1-c_i) y + x-1 = 0,
\end{equation}
leading to
\begin{equation}
    \tau_{i+1} = F^{-1}\left[ \frac {c_i-1+ \sqrt{(c_i+1)^2-4c_ix}}{2c_i} \right],
    \label{eq:tau_next_general}
\end{equation}
where $F^{-1}$ denotes the inverse function of $F(\tau)$. The sign of the square root was chosen to be ``+'' for satisfying $y=1$ for $x=0$ and $y\to 0$ for $x\to 1$.

By our copula-based algorithm, one can generate the event sequence without predetermining the number of IETs as well as with $P(\tau)$ and $M$ being controlled independently. The generating procedure is as depicted in Fig.~\ref{fig:diagram}: We randomly draw the first IET from $P(\tau)$, denoted by $\tau_1$. Then $\tau_i$ is used to generate $\tau_{i+1}$ using $P(\tau_{i+1}|\tau_i)$ or Eq.~\eqref{eq:tau_next_general} with a random number $x$ independently drawn for each $i=1,2,\cdots$. This algorithm can be called Markovian in the sense that $\tau_{i+1}$ depends only on $\tau_i$ but not on $\tau_{i'}$ with $i'<i$.

\section{Results}\label{sec:result}

We apply the copula-based algorithm to three well-known interevent time (IET) distributions, i.e., exponential and power-law IET distributions as well as power-law IET distribution with exponential cutoff. We also test the performance of our algorithm for these three cases, in comparison to the shuffling method~\cite{Hiraoka2018Correlated}.

\subsection{Exponential IET distribution}

\begin{figure}[!t]
    \includegraphics[width=\columnwidth]{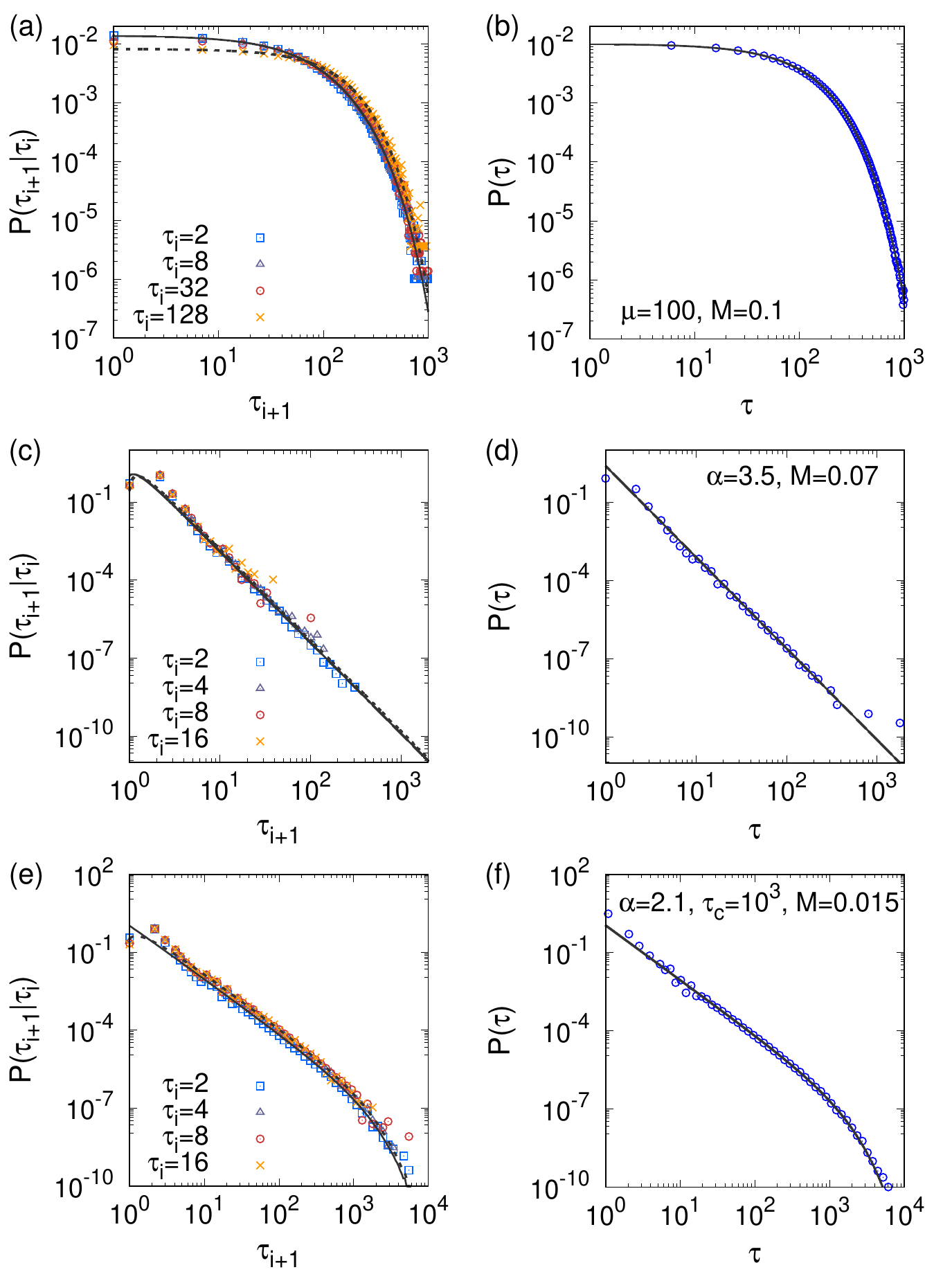}
    \caption{Numerical results of the copula-based algorithm: Conditional probability distributions $P(\tau_{i+1}|\tau_i)$ for several values of $\tau_i$ (left) and the resultant IET distribution (right) for the exponential case in Eq.~\eqref{eq:Ptau_expo} with $\mu=100$ and $M=0.1$ (top), for the power-law case in Eq.~\eqref{eq:Ptau_power} with $\alpha=3.5$ and $M=0.07$ (middle), and for the case of power law with exponential cutoff in Eq.~\eqref{eq:Ptau_cutoff} with $\alpha=2.1$, $\tau_c=10^3$, and $M=0.015$ (bottom) are compared to analytical curves in Eqs.~\eqref{eq:cond_P_expo},~\eqref{eq:Ptau_expo},~\eqref{eq:cond_P_power},~\eqref{eq:Ptau_power},~\eqref{eq:cond_P_cutoff}, and~\eqref{eq:Ptau_cutoff} (from a to f), respectively. In (a), the solid and dashed curves correspond to $\tau_i=2$ and $128$, respectively. In (c,e) the solid and dashed curves correspond to $\tau_i=2$ and $16$, respectively. In (a,b) the curves are linearly binned, while the curves in (c--f) are all log-binned. Numerical results are averaged over $100$ event sequences of size $n=10^5$.}
    \label{fig:copula}
\end{figure}

As a simple example, we consider the exponential IET distribution with the mean IET $\mu$:
\begin{eqnarray}
    \label{eq:Ptau_expo}
    P(\tau) &=& \mu^{-1} e^{-\tau/\mu},\\
    F(\tau) &=& 1- e^{-\tau/\mu},
\end{eqnarray}
from which we get
\begin{equation}
    r = 4M.
\end{equation}
This implies that $|M| \leq 1/4$ by Eq.~\eqref{eq:Mr_ratio}. The conditional probability distribution in Eq.~\eqref{eq:cond_P} is written as
\begin{eqnarray}
    &&P(\tau_{i+1}|\tau_{i}) = \mu^{-1} e^{-\tau_{i+1}/\mu} \nonumber \\
    &&\times \left[ 1+ 4M (1-2 e^{-\tau_i/\mu}) (1-2 e^{-\tau_{i+1}/\mu}) \right].
    \label{eq:cond_P_expo}
\end{eqnarray}
Then from Eq.~\eqref{eq:tau_next_general} we obtain $\tau_{i+1}$ for a given $\tau_i$ and a random number $x$ as
\begin{equation}
    \tau_{i+1} = \mu \ln \left[\frac{2c_i}{c_i+1 - \sqrt{(c_i+1)^2-4c_ix}}\right],
\end{equation}
where 
\begin{equation}
    c_i=rf(\tau_i)= 4M (1-2e^{-\tau_i/\mu}).
\end{equation}

For the demonstration, we generate $100$ event sequences of size $n=10^5$ for $\mu=100$ and $M=0.1$ to analyze them by measuring the conditional probability distributions and the resultant IET distribution, which are comparable to the corresponding analytical curves, as shown in Fig.~\ref{fig:copula}(a,b). We also measure the memory coefficient $M=0.100(4)$ from the same generated event sequences, which is in good agreement with the input value of $0.1$.

\subsection{Power-law IET distribution}

Next, we study the case with a power-law IET distribution with power-law exponent $\alpha$:
\begin{eqnarray}
    \label{eq:Ptau_power}
    P(\tau) &=& (\alpha - 1)\tau^{-\alpha}\theta(\tau-1),\\
    F(\tau) &=& (1 - \tau^{1-\alpha})\theta(\tau-1),
\end{eqnarray}
where $\theta(\cdot)$ denotes the Heaviside step function and the lower bound of IET has been set to be $1$. We assume that $\alpha>3$ for the finite variance of IETs. The relation between $r$ and $M$ is obtained as
\begin{equation}
    r = \frac{(2\alpha-3)^{2}}{(\alpha-1)(\alpha-3)} M.
\end{equation}
The conditional probability distribution in Eq.~\eqref{eq:cond_P} for $\tau_i\geq 1$ and $\tau_{i+1}\geq 1$ is written as 
\begin{eqnarray}
    &&P(\tau_{i+1}|\tau_{i}) = (\alpha - 1)\tau_{i+1}^{-\alpha} \nonumber \\
    &&\times \left[ 1+ \tfrac{(2\alpha-3)^{2}}{(\alpha-1)(\alpha-3)} M (1-2 \tau_i^{1-\alpha}) (1-2 \tau_{i+1}^{1-\alpha}) \right].
    \label{eq:cond_P_power}
\end{eqnarray}
Then from Eq.~\eqref{eq:tau_next_general}, we obtain $\tau_{i+1}$ for a given $\tau_i$ and a random number $x$ as
\begin{equation}
    \tau_{i+1} = \left[\frac{2c_i}{c_i+1 - \sqrt{(c_i+1)^2-4c_ix}}\right]^{1/(\alpha-1)}.
\end{equation}
where 
\begin{equation}
    c_i=rf(\tau_i)= \frac{(2\alpha-3)^{2}}{(\alpha-1)(\alpha-3)} M (1-2\tau_i^{1-\alpha}).
\end{equation}

Using our copula-based algorithm we generate $100$ event sequences of size $n=10^5$ for $\alpha=3.5$ and $M=0.07$. Note that $\alpha=3.5$ leads to $|M|\leq a=5/64\approx 0.078$. From the generated event sequences the conditional probability distributions and the resultant IET distribution are measured and compared to the corresponding analytical curves, as shown in Fig.~\ref{fig:copula}(c,d). We also find the measured memory coefficient $M=0.08(1)$ comparable to the input value of $0.07$.

\subsection{Power-law IET distribution with exponential cutoff}

As a more realistic case evidenced by empirical results~\cite{Karsai2018Bursty}, we consider a power-law IET distribution with exponential cutoff $\tau_c$ as
\begin{eqnarray}
    \label{eq:Ptau_cutoff}
    P(\tau) &=& \frac{\tau_{c}^{\alpha-1}}{\Gamma \left( 1-\alpha, 1/\tau_{c} \right)} \tau^{-\alpha} e^{-\tau/\tau_{c}}\theta(\tau-1),\\
    F(\tau) &=& \left[1 - \frac{\Gamma \left( 1-\alpha, \tau/\tau_{c} \right)} {\Gamma \left( 1-\alpha, 1/\tau_{c} \right)}\right]\theta(\tau-1),
\end{eqnarray}
where $\Gamma(\cdot,\cdot)$ is the upper incomplete Gamma function and the lower bound of IET has been set to be $1$. Note that the IET distribution in Eq.~\eqref{eq:Ptau_cutoff} reduces to that in Eq.~\eqref{eq:Ptau_power} in the limit of $\tau_c\to\infty$, or to that in Eq.~\eqref{eq:Ptau_expo} if $\alpha=0$ and the lower bound of IET is set to be $0$. In contrast to the exponential and power-law cases, in the case of power law with exponential cutoff the analytic calculation of the integration in Eq.~\eqref{eq:Mr_ratio} is not straightforward, hence $a$ in Eq.~\eqref{eq:Mr_ratio} will be numerically evaluated. With this numerical value of $a$, the conditional probability distribution in Eq.~\eqref{eq:cond_P} is written as 
\begin{eqnarray}
    &&P(\tau_{i+1}|\tau_{i}) = \tfrac{\tau_{c}^{\alpha-1}}{\Gamma \left( 1-\alpha, 1/\tau_{c} \right)} \tau_{i+1}^{-\alpha} e^{-\tau_{i+1}/\tau_{c}} \nonumber \\
    &&\times \left\{ 1+ \frac{M}{a}
        \left[1 - 2\tfrac{\Gamma \left( 1-\alpha, \tau_{i}/\tau_{c} \right)} {\Gamma \left( 1-\alpha, 1/\tau_{c} \right)}\right]
        \left[1 - 2\tfrac{\Gamma \left( 1-\alpha, \tau_{i+1}/\tau_{c} \right)} {\Gamma \left( 1-\alpha, 1/\tau_{c} \right)}\right]
    \right\}.\nonumber\\
    \label{eq:cond_P_cutoff}
\end{eqnarray}
Then the value of $\tau_{i+1}$ can be obtained for a given $\tau_i\geq 1$ and a random number $x$ by numerically solving Eq.~\eqref{eq:tau_next_general}, namely,
\begin{equation}
    \frac{\Gamma(1-\alpha,\tau_{i+1}/\tau_c)}{\Gamma(1-\alpha,1/\tau_c)}= \frac{c_i+1 - \sqrt{(c_i+1)^2-4c_ix}}{2c_i},
    \label{eq:equation_cutoff}
\end{equation}
where 
\begin{equation}
    c_i=rf(\tau_i)= \frac{M}{a} \left[1 - 2\frac{\Gamma \left( 1-\alpha, \tau_i/\tau_{c} \right)}{\Gamma \left( 1-\alpha, 1/\tau_{c} \right)}\right].
    \label{eq:equation_cutoff_c}
\end{equation}

We generate $100$ event sequences of size $n=10^5$ for $\alpha=2.1$, $\tau_c=10^3$, and $M=0.015$. Note that for $\alpha=2.1$ and $\tau_c=10^3$, we have $|M|\leq a \approx 0.020$. The generated event sequences are analyzed to result in the conditional probability distributions and the resultant IET distribution that are comparable to the corresponding analytical curves, as shown in Fig.~\ref{fig:copula}(e,f). The measured memory coefficient $M=0.015(5)$ is found to be close to the input value of $0.015$. All these results indicate that the correlations between two consecutive IETs are successfully implemented.

\subsection{Computation times}\label{subsec:performance}

We discuss the performance of our algorithm in terms of computation times for generating event sequences with correlated IETs. For the exponential and power-law IET distributions, the generation of event sequences is easy and fast, while for the power-law IET distribution with exponential cutoff (``power+cutoff'' in short) one needs to numerically solve Eq.~\eqref{eq:equation_cutoff} for the generation of each IET, implying longer computation times than other cases. To study this issue, we measure the average computation times in seconds for generating event sequences. We use codes written in C on a Linux system with 3.5 GHz Intel Core i5-7600 CPU and 16 GB RAM. We also use GNU Scientific Library for calculating the incomplete Gamma function, e.g., in Eqs.~\eqref{eq:equation_cutoff} and~\eqref{eq:equation_cutoff_c}. 

\begin{figure}[!t]
    \includegraphics[width=\columnwidth]{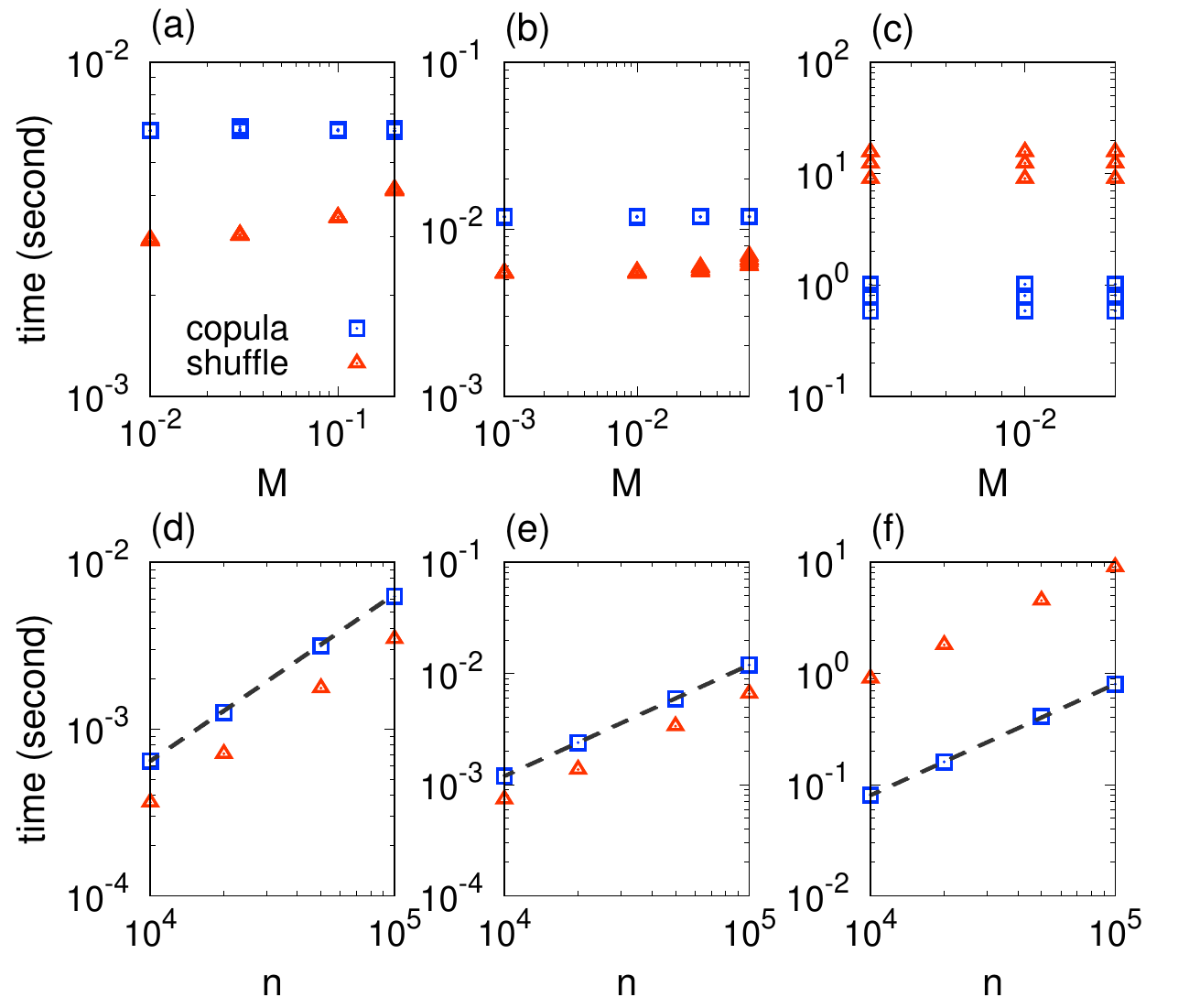}
    \caption{Average computation times in seconds for generating an event sequence using the copula-based algorithm (square) and the shuffling method (triangle) for the exponential case (a,~d), the power-law case (b,~e), and the case of power law with exponential cutoff (``power+cutoff'' in short) (c,~f). In panels (a--c), the computation times for various values of parameters (see the text) are plotted as a function of $M$ for $n=10^5$. In panels (d--f), we show the $n$-dependence of computation times when the event sequences are generated using the same parameter values as in Fig.~\ref{fig:copula}, where the dashed line of $y=x$ is for guiding the eye. Each point is averaged over $100$ event sequences for exponential and power-law cases and $10$ event sequences for the power+cutoff case.}
    \label{fig:time}
\end{figure}

We test various combinations of parameter values for the estimation of computation times when $n=10^5$ is fixed. For the exponential case, we use $\mu=10, 10^2, 10^3$ and $M=0.01,0.03,0.1,0.2$. For the power-law case, we use $\alpha=3.3,3.5,4$ and $M=0.001,0.01,0.03,0.07$. For the power+cutoff case, $\alpha=2.1,2.5,2.9$ and $M=0.005,0.01,0.015$ are used for a fixed $\tau_c=10^3$. Figure~\ref{fig:time}(a--c) shows that the average computation times in the power+cutoff case are larger than those for other two cases, as expected. We observe that the computation times are linearly increasing with $n$ in Fig.~\ref{fig:time}(d--f), which is trivial for the copula-based algorithm.

\begin{figure}[!t]
    \includegraphics[width=\columnwidth]{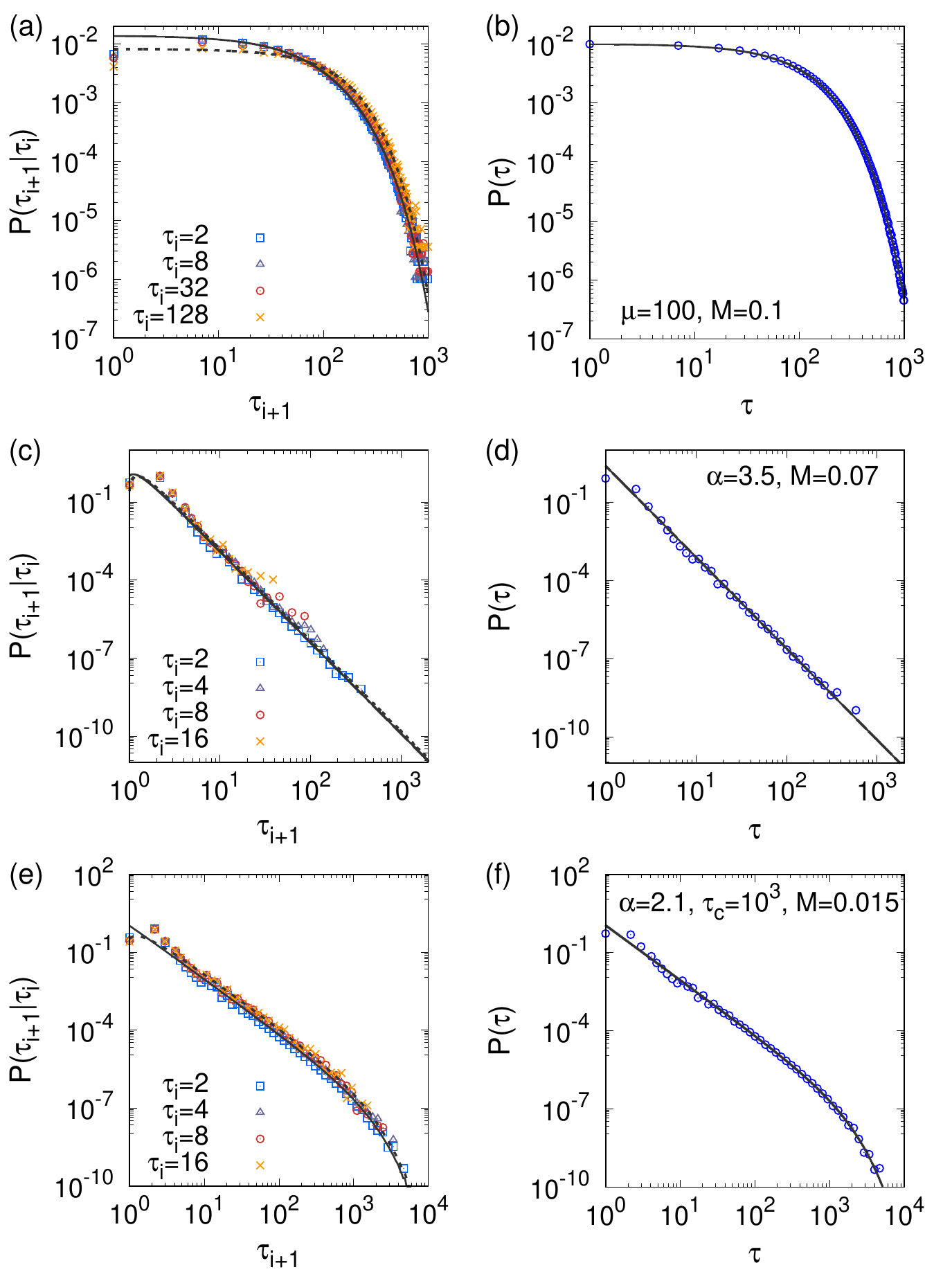}
    \caption{Numerical results of the shuffling method with the same parameter values as in Fig.~\ref{fig:copula}, with $\epsilon=10^{-3}$.}
    \label{fig:shuffle}
\end{figure}

We now compare the computation times of the copula-based algorithm to those of the shuffling method in Ref.~\cite{Hiraoka2018Correlated}. For implementing the shuffling method, we first draw $n$ random values from $P(\tau)$ to make an IET sequence $T\equiv \{\tau_i\}_{i=1,\cdots,n}$. Using the definition of Eq.~\eqref{eq:memory_original}, we measure the memory coefficient from $T$, denoted by $\tilde M$. Two IETs are randomly chosen in $T$ and swapped only when this swapping makes $\tilde M$ closer to $M$, i.e., the target value. By repeating the swapping, we can obtain the IET sequence whose $\tilde M$ is close enough to $M$. Precisely, the swapping stops when $|M-\tilde M|<\epsilon$ with a small number $\epsilon$.

The computation times for generating event sequences by the shuffling method are estimated for the same combinations of parameter values as in the copula-based algorithm, together with several values of $\epsilon=0.001,\cdots, 0.02$. Note that the computation time for the shuffling method is the sum of the time for generating the initial set of IETs for each given IET distribution and the time for the shuffling procedure. In Fig.~\ref{fig:time}(a--c) we find that the copula-based algorithm outperforms the shuffling method only for the power+cutoff case by a factor of $10$, while for the exponential and power-law cases the shuffling method is around twice faster than the copula-based algorithm. Therefore, the shuffling procedure itself seems relatively fast. Then much slower generation by the shuffling method in the power+cutoff case could be due to the longer time for generating the initial set of IETs using Eq.~\eqref{eq:Ptau_cutoff}~\footnote{To generate the IETs from the power-law distribution with exponential cutoff, we have used the method described in Ref.~\cite{Clauset2009Powerlaw}, which indeed requires one more random number per each IET than the cases with exponential and power-law IET distributions.}. We also find that the computation times for the shuffling method are mostly insensitive to the variation of parameter values, except for the effect of $M$ in the exponential and power-law cases. This can be explained by the fact that the initial sequence of IETs drawn from $P(\tau)$ is uncorrelated, i.e., $\tilde M\approx 0$, implying that it takes longer time to reach the larger target value of $M$. Such $M$-dependence is not apparent in the power+cutoff case probably due to the dominant effects of the time for generating the initial set of IETs. In addition, the computation times turn out to be linearly increasing with $n$ in Fig.~\ref{fig:time}(d--f) as in the copula-based algorithm.

Finally the quality of event sequences generated by the shuffling method is tested by measuring the conditional probability distribution $P(\tau_{i+1}|\tau_i)$ and the IET distribution $P(\tau)$ from the generated event sequences using the same parameter values as in Fig.~\ref{fig:copula}. Here we choose $\epsilon=10^{-3}$ which is of the order of the standard deviations of estimated $M$ in the case with the copula-based algorithm. In Fig.~\ref{fig:shuffle} we find almost the same behaviors as in the results from the copula-based algorithm, implying that the copula-based algorithm and the shuffling method can be used interchangeably. It might be due to the fact that both methods are designed to implement only the correlations between two consecutive IETs for a given IET distribution, while randomizing or ignoring all other higher-order correlations between IETs. In this sense, another generative method called the Laplace Gillespie algorithm (LGA)~\cite{Masuda2018Gillespie} can also be compared to our algorithm and shuffling method as it can generate event sequences with correlated IETs. However, we leave this comparison as a future work mainly because fine-tuning the value of $M$ by the LGA seems to be difficult.

\section{Conclusion}\label{sec:conclusion}

We have proposed the Farlie-Gumbel-Morgenstern (FGM) copula-based algorithm for generating event sequences with correlated interevent times (IETs) for a given IET distribution and a given memory coefficient between two consecutive IETs. This is to overcome the disadvantages in the previous generative methods, i.e., the shuffling method in Ref.~\cite{Hiraoka2018Correlated} and the method using conditional probability distribution in Ref.~\cite{Artime2017Dynamics}: By adopting the conditional probability distribution $P(\tau_{i+1}|\tau_i)$ based on the FGM copula~\cite{Takeuchi2010Constructing, Nelsen2006Introduction}, one can generate an arbitrary number of IETs without predetermining the number of IETs, while the statistics of IETs and their correlations can be independently controlled. After deriving the analytical forms of the next IET $\tau_{i+1}$ for a given previous IET $\tau_i$, we show that our algorithm successfully generates the event sequences with desired statistical properties.

We also compare the performance of our copula-based algorithm to the shuffling method in terms of computation times: It turns out that the copula-based algorithm outperforms the shuffling method for generating event sequences with power-law IET distributions with exponential cutoff, while for the exponential and power-law IET distributions the shuffling method is around twice faster than the copula-based algorithm. We note that since both methods generate event sequences of the same statistical properties, any of them can be used appropriately. Considering the advantages of our algorithm, we expect our algorithm to be used for modeling and simulating more realistic event sequences, eventually for more realistic temporal networks.

Finally we remark on the limited range of $M$. Apart from the bounds of $M$ set by the shape of $P(\tau)$~\cite{Guo2017Bounds}, the range of $M$ is also bounded due to the form of the FGM copula, while it is more flexible in the shuffling method. Therefore, other members of the FGM family, e.g., the iterated FGM copula~\cite{Huang1984Correlation}, can be investigated to explore a wider range of $M$ for the IET distributions of our interest as a future work. In addition, our copula-based algorithm can be used to generate any other sequence of correlated variables.

\begin{acknowledgments}
    H.-H.J. was supported by Basic Science Research Program through the National Research Foundation of Korea (NRF) funded by the Ministry of Education (NRF-2018R1D1A1A09081919).
    W.-S.J. was supported by Basic Science Research Program through the National Research Foundation of Korea (NRF) funded by the Ministry of Education (2016R1D1A1B03932590).
\end{acknowledgments}

\bibliographystyle{apsrev4-1}
%
  
\end{document}